\documentclass{article}
\usepackage{spconf,amsmath,graphicx,hyperref}
\usepackage[whole]{bxcjkjatype}
\usepackage{colortbl,xcolor}
\usepackage{amssymb}

\title{Paralinguistic Emotion-Aware Validation Timing Detection \\in Japanese Empathetic Spoken Dialogue}
\name{Zi Haur Pang, Yahui Fu, Yuan Gao, Tatsuya Kawahara}
\address{Graduate School of Informatics, Kyoto University, Kyoto, Japan\\
\texttt{\{pang, fu, gao, kawahara\}@sap.ist.i.kyoto-u.ac.jp}}
\begin{document}
\ninept
\maketitle
\begin{abstract}
Emotional Validation is a psychotherapy communication technique that involves recognizing, understanding, and explicitly acknowledging another person's feelings and actions, which strengthens alliance and reduces negative affect. To maximize the emotional support provided by validation, it is crucial to deliver it with appropriate timing and frequency. This study investigates validation timing detection from the speech perspective. Leveraging both paralinguistic and emotional information, we propose a paralinguistic- and emotion-aware model for validation timing detection without relying on textual context. Specifically, we first conduct continued self-supervised training and fine-tuning on different HuBERT backbones to obtain (i) a paralinguistics-aware Self-Supervised Learning (SSL) encoder and (ii) a multi-task speech emotion classification encoder. We then fuse these encoders and further fine-tune the combined model on the downstream validation timing detection task. Experimental evaluations on the TUT Emotional Storytelling Corpus (TESC) compare multiple models, fusion mechanisms, and training strategies, and demonstrate that the proposed approach achieves significant improvements over conventional speech baselines. Our results indicate that non-linguistic speech cues, when integrated with affect-related representations, carry sufficient signal to decide when validation should be expressed, offering a speech-first pathway toward more empathetic human–robot interaction.
\end{abstract}
\begin{keywords}
Emotional Validation, Empathy, Spoken Dialogue Systems (SDSs)
\end{keywords}
\section{Introduction}
\label{sec:intro}

Empathy is central to human-robot and human-agent interaction: when systems show that they ``get'' the user, people disclose more, follow guidance, and build trust. In counseling and mental-health contexts, empathic virtual agents and robots have been linked to higher perceived empathy, trust, and satisfaction, while cautioning that poorly timed or generic displays can feel inauthentic \cite{szondy2024attachment, leite2013influence}. Yet common empathic phrases (e.g., \textit{I'm sorry to hear that}) often sound formulaic, especially for users who suppress or struggle to name emotions. In psychotherapy, a more specific technique, \textbf{emotional validation (validation)}, helps. Validation means recognizing, understanding, and explicitly acknowledging another person's feelings and actions; classic work formalized multiple levels of validation and tied the skill to better alliance and reduced negative affect \cite{linehan1997validation}. Subsequent studies report that validation predicts reductions in distress and better engagement across clinical and youth settings \cite{wasson2022youths,carson2018effect}. 

Crucially, everyday validation arrives through both words and paralinguistic signals: tone, pitch, loudness, timing, and short listener responses (backchannels). Speech science shows that vocal cues carry emotion reliably and shape judgments such as warmth and credibility, even when lexical content is held constant \cite{banse1996acoustic,guyer2021paralinguistic}. A practical challenge follows: when should a system validate? Over-validating can feel insincere; under-validating misses support. Conversation research indicates that the timing of listener responses is guided by acoustic--prosodic patterns and local interactional signals. In Japanese conversation, short listener responses (aizuchi) occur frequently and at specific positions, with prosodic triggers such as low-pitch regions, pauses, and pre-boundary lengthening; these cues can invite backchannels or signal imminent turn-yielding \cite{itzchakov2023listening, kraus2017voice}. These findings suggest that appropriate validation timing may be detectable from speech alone, without relying on lexical context.

In dialogue systems, empathy has been pursued via commonsense reasoning \cite{sabour2022cem}, emotion-cause extraction \cite{gao2021improving}, and persona conditioning \cite{fu2024styemp, cai2024pecer}. However, only a few works target \emph{validation} directly, and most are text-based. Recent work frames validation as a pipeline, from validation timing detection, emotion identification, to validating response generation, with initial results in Japanese empathetic dialogue, but it remains open how far non-linguistic speech cues alone can carry the signal \cite{pang2023prediction,pang2024acknowledgment}. At the same time, there is a rich line of speech research on turn-taking and backchannel prediction that leverages prosody and timing for listener behavior, including Japanese conversations and multi-modal settings, suggesting a promising foundation for speech-first validation timing \cite{ward2000prosodic,gravano2011turn}.

This paper tests the hypothesis that correct validation timing can be detected from speech alone by jointly modeling paralinguistic and emotional information in Japanese emotional dialogue. We isolate the timing decision (``validate now or not'') without using lexical context and propose a two-branch speech approach: (i) continued pre-training of a HuBERT encoder on emotional dialogue to emphasize paralinguistic patterns, and (ii) multi-task fine-tuning of a second HuBERT encoder to learn emotion cues. We then fuse both encoders and fine-tune for the final validation timing detector. Using the TUT Emotional Storytelling Corpus (TESC) for evaluation, we comprehensively compare against strong speech and language baselines and analyze fusion and training strategies.

\begin{figure*}[t]
    \centering
\includegraphics[width=0.8\linewidth]{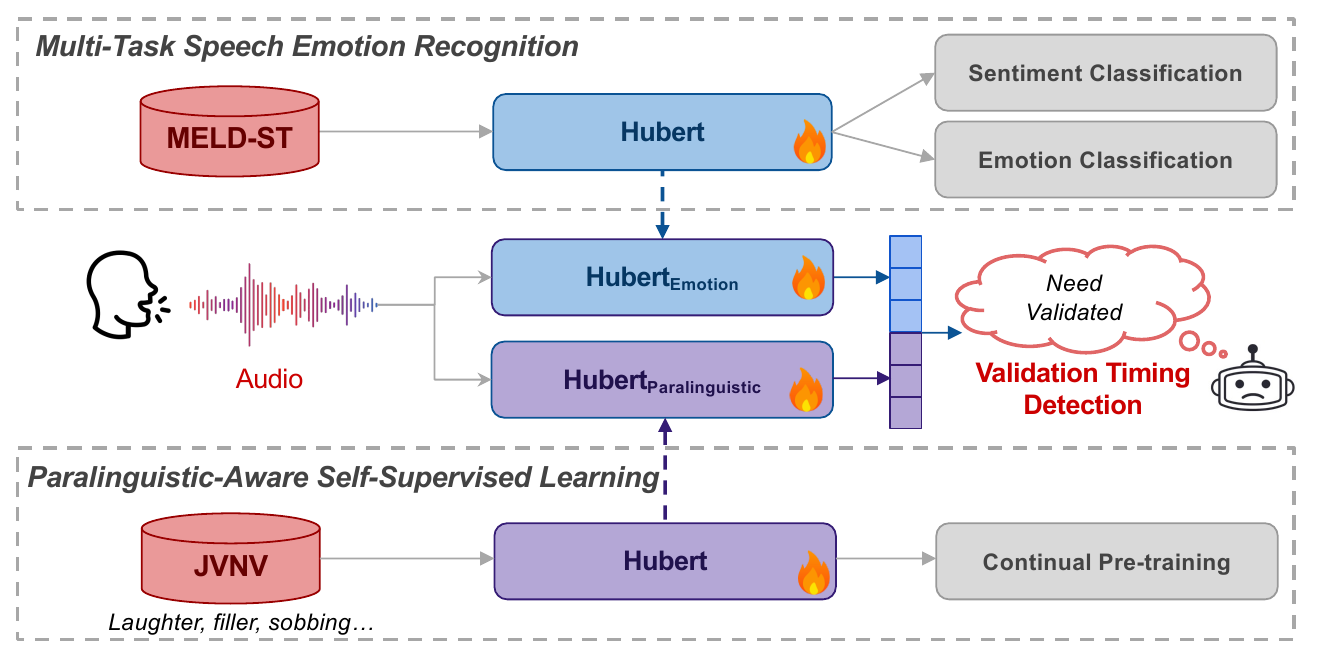}
    \caption{Architecture of the proposed method. The method consists of (1) a multitask speech emotion recognition model, (2) a paralinguistic-aware self-supervised learning model, and (3) feature fusion for the validation timing detection task.}
    \label{fig:arch}
\end{figure*}


\section{Proposed Methods}

\subsection{Multi-Task Speech Emotion Recognition}
\label{sec:multitask}

Current self-supervised speech encoders such as HuBERT and wav2vec~2.0 are pre-trained on general speech and may therefore mismatch our downstream task on emotional spoken dialogue. To reduce this domain gap, we fine-tune a pre-trained model and enrich feature learning with auxiliary tasks under a multi-task learning (MTL) framework. As shown in top part of Fig.~\ref{fig:arch}, given an input waveform $U$ we take the last-layer hidden sequence $\mathbf{f}\!\in\!\mathbb{R}^{t\times d}$ from the backbone as the latent representation, where $t$ is the utterance length and $d$ the hidden dimension. We apply mean pooling over time (excluding padded frames) to obtain an utterance-level vector
\begin{equation}
\bar{\mathbf{h}}=\frac{1}{t}\sum_{i=1}^{t}\mathbf{f}_{i}\in\mathbb{R}^{d},
\end{equation}
followed by dropout ($p{=}0.1$). Two linear heads map $\bar{\mathbf{h}}$ to emotion and sentiment logits, respectively:
\begin{align}
\mathbf{z}^{e} &= \mathbf{W}^{e}\bar{\mathbf{h}}+\mathbf{b}^{e}, \quad
\mathbf{p}^{e}=\mathrm{softmax}(\mathbf{z}^{e}), \\
\mathbf{z}^{s} &= \mathbf{W}^{s}\bar{\mathbf{h}}+\mathbf{b}^{s}, \quad
\mathbf{p}^{s}=\mathrm{softmax}(\mathbf{z}^{s}).
\end{align}

In this MTL branch, \emph{emotion} is modeled as a fine-grained 7-way categorical prediction (\emph{anger, disgust, fear, joy, neutral, sadness, surprise}), while \emph{sentiment} is modeled as a coarse 3-way polarity prediction
(\emph{negative, neutral, positive}). 
Learning both encourages complementary affect representations: the emotion head captures discrete affect categories, whereas the sentiment head provides an auxiliary polarity signal that regularizes the shared utterance representation.

To mitigate label skew in emotion classification, we use a class-weighted cross-entropy; sentiment uses standard cross-entropy:
\begin{align}
\mathcal{L}_{\mathrm{emotion}} &=
-\!\!\sum_{c=1}^{C_e} w^{(e)}_c\, y^{e}_c \log p^{e}_c, \\
\mathcal{L}_{\mathrm{sentiment}} &=
-\!\!\sum_{c=1}^{C_s} y^{s}_c \log p^{s}_c,
\end{align}
where $C_e{=}7$ and $C_s{=}3$, $y^{e},y^{s}$ are one-hot targets, and $w^{(e)}_c$ are inverse-frequency weights estimated on the training split.

Instead of fixing the task weights, we learn them end-to-end. Specifically, a scalar logit parameter $\tilde{\alpha}$ is optimized jointly with model parameters and mapped through a sigmoid, $\alpha=\sigma(\tilde{\alpha})\in(0,1)$. The total loss is
\begin{equation}
\mathcal{L}=\alpha\,\mathcal{L}_{\mathrm{emotion}}+(1-\alpha)\,\mathcal{L}_{\mathrm{sentiment}},
\end{equation}
which adapts the balance between tasks during training and removes manual tuning. We initialize $\tilde{\alpha}{=}0$ so that $\alpha{=}0.5$ at the start of training.

\subsection{Paralinguistic-Aware Self-Supervised Learning}
\label{sec:CPT}

Motivated by psychological evidence linking paralinguistics and validation \cite{guyer2021paralinguistic}, we conduct continue pre-training with a HuBERT-style masked unit prediction objective \cite{hsu2021hubert}, as in Fig.~\ref{fig:arch} bottom. We first derive discrete pseudo-target units by running $k$-means on MFCC frames and align the resulting unit IDs to the encoder frame rate. Let $K$ be the number of clusters and $\{\mathbf{h}_t\}_{t=1}^{T}$ the frame-level hidden states from the encoder. A linear classifier maps each state to unit logits $\mathbf{z}_t\!\in\!\mathbb{R}^{K}$:
\begin{equation}
\mathbf{z}_t=\mathbf{W}\mathbf{h}_t+\mathbf{b}, \qquad u_t\in\{1,\dots,K\}.
\end{equation}

We apply time-span masking on the feature axis by sampling span starts and masking consecutive steps; only masked, non-padded positions contribute to the loss. Denoting by $\mathcal{A}$ the set of masked, non-padded indices,
\begin{equation}
\mathcal{L}_{\text{SSL}}
=\frac{1}{|\mathcal{A}|}\sum_{t\in\mathcal{A}}
\mathrm{CE}\!\left(\mathbf{z}_t, u_t\right).
\end{equation}
Feature-rate lengths are computed from raw attention masks via the backbone's length mapping, and feature-rate mask indices are generated accordingly. Labels on padded steps are set to excluded from the loss. The resulting encoder provides a paralinguistics-aware representation used as one branch in our downstream fusion.

\subsection{Feature Fusion}

To detect the correct timing to express validation, we concate the two encoders introduced in Sections~\ref{sec:multitask} and~\ref{sec:CPT}, as shown in middle of Fig.~\ref{fig:arch}. Each encoder produces a frame-level hidden-state sequence; we obtain utterance embeddings by mean pooling over unpadded frames, with frame counts derived from each encoder's feature extractor to avoid padding effects.

Both embeddings are projected into a shared space (projection dimension $256$) using linear layers followed by GELU and dropout $0.1$. We then concatenate the projected embeddings and apply a final linear layer to produce two-way logits for validate or not, trained with cross-entropy.

\section{Experimental Setup}

\subsection{Dataset}

\subsubsection{Pre-training Dataset}

For multi-task speech emotion recognition pre-training, we use MELD-ST \cite{chen2024meld}, a multilingual extension of MELD \cite{poria2019meld} that provides English$\rightarrow$Japanese (En–Ja) and English$\rightarrow$German (En–De) pairs. Each pair contains about 10{,}000 utterances with emotion labels, extracted from the TV series ``Friends" and rendered as acted speech by professional speakers. In this study, we use only the Japanese portion and re-split the corpus to remove speaker overlap while matching split sizes and label distributions, yielding Train/Valid/Test ratios of $0.71/0.13/0.16$ with speakers $199/36/46$ and utterances $6665/1623/1797$.

For paralinguistic-aware self-supervised pre-training, we use the JVNV corpus \cite{xin2024jvnv}, a Japanese emotional speech dataset that includes both verbal content and nonverbal vocalizations. JVNV comprises approximately 3.94~hours of acted speech from four voice actors (two male, two female) across six emotions (anger, disgust, fear, happiness, sadness, surprise). Each utterance expresses a single target emotion. We create a speaker-disjoint split into train and validation sets with an 8:2 ratio.

\subsubsection{Validation Timing Detection Dataset}
We evaluate on the TUT Emotional Storytelling Corpus (TESC) \cite{oishi2021design}, a Japanese two-party, multi-turn spoken-dialogue dataset where close friends share personal experiences prompted by one of eight Plutchik emotions \cite{plutchik2001nature}. TESC contains 247 sessions from 18 pairs (about 9.2 hours total; 133.9 s/session on average). Using the audio with supporting ASR transcripts, annotators label each utterance end as \textit{validate} (a validating response should follow) or \textit{non-validate}. We split the annotated data into train/validation/test with a ratio of 7:1.5:1.5; reported in Table~\ref{tab:data_composition}.

\begin{table}[t]
\caption{Distribution of the dataset}
\centering
\small
\begin{tabular}{lcc}
\hline
Dataset & \#Validation & \#Non-Validation \\
\hline
TESC & 489 & 881 \\
\textit{-train} & 340 & 579 \\
\textit{-val} & 74 & 157 \\
\textit{-test} & 75 & 145 \\
\hline
\end{tabular}

\label{tab:data_composition}
\end{table}

\subsection{Implementation Details}

\subsubsection{Multi-Task Speech Emotion Recognition}

Using Japanese HuBERT-Large\footnotemark[1] as the backbone, we fine-tune with AdamW \cite{loshchilovdecoupled} at a learning rate of $1{\times}10^{-5}$, an effective batch size of 16 via gradient accumulation, and early stopping (patience 5) on the validation set. We evaluate every 500 steps and select the checkpoint that maximizes macro-F1. The resulting model attains emotion recognition unweighted/weighted accuracy of $30.43/46.13$ and sentiment unweighted/weighted accuracy of $52.06/54.81$.

\subsubsection{Paralinguistic-Aware Self-Supervised Learning}

For continued pre-training, we use Japanese HuBERT-Large\footnotemark[1] with $K{=}100$ $k$-means MFCC units (fit on 200k frames) and span masking ($p{=}0.065$, length $10$). We train for 20 epochs at learning rate $1{\times}10^{-5}$ with batch size 2 and 16-step gradient accumulation, validate each epoch, and checkpoint the lowest validation loss.

\subsubsection{Validation Timing Detection}

We trained all models on a single NVIDIA RTX~6000 Ada GPU, fine-tuning on TESC with learning rate $1{\times}10^{-5}$, batch size 1 with 16-step gradient accumulation (max 20 epochs), 100-step warmup, and weight decay 0.01. We validated every 100 steps, used early stopping (patience 5), selected the best checkpoint by F1, and fixed the random seed to 42. To mitigate the validate/non-validate skew (340 vs.\ 579) on this limited-scale dataset, we balanced the training split by downsampling non-validation utterances with 250 to 329.

\subsection{Evaluation Metrics}

In everyday conversation, when a system chooses to validate matters more than how often it does so. Accordingly, we treat target-class precision, which is the proportion of predicted \textit{validate} turns that truly warrant validation, as the principal metric. A model that labels many turns as \textit{validate} (high recall) risks hollow or repetitive acknowledgments that undermine perceived empathy; thus a high F1 score alone can be misleading if it masks low precision. We therefore report (i) \textbf{validation precision (V-Pre.)} as the primary indicator of conversational appropriateness, (ii) \textbf{validation F1 (V-F1)} to capture the precision–recall trade-off on the target class, (iii) \textbf{non-validation F1 (NV-F1)} to capture the trade-off on the non-target class, and (iv) \textbf{macro-averaged F1 (M-F1)} across both classes to ensure performance on the majority \textit{non-validate} class is not overlooked.

\section{Results and Analysis}

\subsection{Results of Validation Timing Detection}

We first evaluate our method against publicly available speech models, namely Japanese HuBERT-Large\footnote{\url{https://huggingface.co/rinna/japanese-hubert-large}} and xlsr-53-japanese\footnote{\url{https://huggingface.co/jonatasgrosman/wav2vec2-large-xlsr-53-japanese}}. The results in Table~\ref{tab:model_performance} show that both baselines perform poorly on this task, likely because their pre-training targets general acoustic regularities rather than emotional dialogue, creating a substantial domain mismatch. In contrast, our method, which combines emotion-aware and paralinguistic cues, successfully identifies appropriate moments to validate, achieving a validation precision of $47.96$ and a validation F1 of $54.34$, while maintaining a strong macro F1 of $62.37$.

Beyond speech encoders, because our main objective is to test whether we can identify the just-right timing to express validation without understanding textual context, we also compare against publicly available language models: traditional language model (LMs) such as BERT\footnote{\url{https://huggingface.co/tohoku-nlp/bert-large-japanese}} and ModernBERT\footnote{\url{https://huggingface.co/sbintuitions/modernbert-ja-130m}}, as well as large language models (LLMs) such as Llama~3.1~8B~Instruct\footnote{\url{https://huggingface.co/meta-llama/Llama-3.1-8B-Instruct}} and GPT\mbox{-}4.1~Nano\footnote{\url{https://openai.com/index/gpt-4-1/}}. For the traditional LMs, we fine-tune directly on our dataset using only transcripts; for LLMs, we evaluate both zero-shot and 3-shot settings. The results indicate that current LLMs (in both zero-shot and 3-shot) struggle to identify correct validation timing, whereas fine-tuned traditional LMs perform better but still lag behind our speech-only approach. Notably, our method surpasses all language-model baselines despite not using textual context, supporting the feasibility of speech-first validation timing in spoken dialogue scenarios where a separate transcription step is unnecessary.


%
\begin{table}[t]
\caption{Results of the validation timing detection task [\%].}
\label{tab:model_performance}
\centering
\begin{tabular}{lcccc}
\hline
\multicolumn{1}{c}{} & V-Prec. & V-F1 & NV-F1 & M-F1 \\ \hline
\rowcolor[gray]{0.95}\textbf{Language Model}  &  & & & \\
BERT  & 39.96 & 41.62 & 66.76 & 54.19 \\
ModernBERT  & 41.32 & 44.65 & 68.65 & 56.65 \\
Llama 3.1 8B  &  & & & \\
\textit{- Zero-shot}  & 33.91 & 49.04 & 40.48 & 44.76 \\
\textit{- 3-shot}  & 33.69 & 50.30 & 46.10 & 48.20 \\
GPT 4.1 Nano  &  & & & \\
\textit{- Zero-shot}  & 39.97 & 49.43 & 58.97 & 54.20 \\
\textit{- 3-shot}  & 39.44 & 46.55 & 61.93 & 54.24 \\
\rowcolor[gray]{0.95}\textbf{Speech Model}  &  & & & \\
HuBERT  & 38.51 & 51.12 & 49.77 & 50.45 \\
Xlsr-53  & 37.04 & 47.62 & 52.17 & 48.89 \\
Proposed (Ours) & \textbf{47.96} & \textbf{54.34} & \textbf{70.41} & \textbf{62.37} \\
\hline
\end{tabular}
\end{table}

\subsection{Ablation Study}

To assess the contribution of each component, we compare: \emph{Para. HuBERT} (the continued pre-trained HuBERT that learns paralinguistic cues from Section~\ref{sec:CPT}), and \emph{Emo. HuBERT} (the standalone emotion encoder from Section~\ref{sec:multitask}). The results in Table~\ref{tab:ablation} indicate that the emotion encoder improves timing detection (validation F1 $52.81$, macro F1 $60.73$), underscoring the relevance of emotional information for timing validation, consistent with text-based findings \cite{pang2023prediction}. Notably, although MELD-ST and JVNV are acted corpora, they still help modeling on the spontaneous TESC corpus. The paralinguistic branch, \emph{Para. HuBERT} attains validation precision $52.54\%$, showing that paralinguistic cues can support validation timing, in line with psychological evidence \cite{guyer2021paralinguistic, weiste2014prosody}.



%
\begin{table}[t]
\caption{Ablation study [\%].}
\label{tab:ablation}
\centering
\begin{tabular}{lcccc}
\hline
\multicolumn{1}{c}{} & V-Prec. & V-F1 & NV-F1 & M-F1 \\ \hline
Para. HuBERT & \textbf{52.54} & 46.27 & \textbf{76.47} & 61.37 \\
Emo. HuBERT & 45.63 & 52.81 & 67.94 & 60.73 \\
Proposed (Ours) & 47.96 & \textbf{54.34} & 70.41 & \textbf{62.37} \\
\hline
\end{tabular}
\end{table}

\subsection{Evaluation of Different Fusion Strategies}

We next evaluate how different fusion strategies affect validation timing detection. Results are summarized in Table~\ref{tab:fusion}. In addition to our proposed simple concatenation method, we test three learned fusion variants: (i) \textbf{Attention}: each encoder embedding $\mathbf{h}\in\mathbb{R}^{d}$ is scored by a two-layer MLP with $tanh$, the scores are normalized across branches, and a weighted sum forms the fused vector (classifier on $\mathbb{R}^{d}$); (ii) \textbf{Gated}: the two projected embeddings are concatenated and passed through a sigmoid gate $\mathbf{g}\in(0,1)^{d}$ to produce an element-wise blend $\mathbf{g}\!\odot\!\mathbf{h}_{1}+(1-\mathbf{g})\!\odot\!\mathbf{h}_{2}$ (classifier on $\mathbb{R}^{d}$); and (iii) \textbf{Multi-Head Attention (MHA)}: the two embeddings are treated as a length-2 sequence and passed through a single multi-head attention layer (4 heads), with the output pooled and fed to a classifier (on $\mathbb{R}^{d}$). Compared with these, \textbf{Concat} (simple feature concatenation in our proposed method) achieves the best validation precision, validation F1, and macro F1 ($47.96\%$, $54.34\%$, $62.37\%$). We attribute this to our small, imbalanced dataset: higher-capacity fusion (attention/gating/MHA) is more prone to overfitting, whereas concatenation is a strong and stable fusion method in low-data settings.

\begin{table}[t]
\caption{Evaluation of different fusion strategies [\%].}
\label{tab:fusion}
\centering
\begin{tabular}{lcccc}
\hline
\multicolumn{1}{c}{} & V-Prec. & V-F1 & NV-F1 & M-F1 \\ \hline
Attention & 43.81 & 51.11 & 66.15 & 58.63 \\
Gated & 44.00 & 50.29 & 67.17 & 58.73 \\
MHA & 46.67 & 50.91 & \textbf{70.55} & 60.73 \\
Concat (Ours) & \textbf{47.96} & \textbf{54.34} & 70.41 & \textbf{62.37} \\
\hline
\end{tabular}
\end{table}

\subsection{Evaluation of Different Training Strategies}

We assess how different training strategies affect validation timing performance; results are summarized in Table~\ref{tab:strategy}. We compare four base settings: (i) freezing both encoders, (ii) fine-tuning either encoder while freezing the another encoder, (iii) fully fine-tuning either encoder while parameter-efficient updates with LoRA on the another encoder, and (iv) fine-tuning both encoders (default).

Fine-tuning both encoders yields the best overall result (V-F1 $54.34$, M-F1 $62.37$), indicating that joint adaptation lets the two branches co-specialize to the target label distribution and complementary cues. Freezing both encoders collapses target-class performance (zeros on V-Prec./V-F1), consistent with frozen-feature limits on specialized timing tasks. Freezing the paralinguistic branch while fine-tuning only the emotion encoder produces the highest precision (V-Prec.\ $61.90$) but very low V-F1 ($27.08$), suggesting an over-conservative boundary with poor recall due to misaligned fixed paralinguistic features. LoRA on one branch (while fine-tuning the other) attains competitive macro scores (M-F1 $60.11$ / $60.67$) but lower V-F1 ($46.15$ / $47.62$), implying reduced capacity for tight co-adaptation compared to full fine-tuning.

\begin{table}[t]
\caption{Evaluation of different training strategies [\%]. Para.: Paralinguistic Encoder; Emo.: Emotion Encoder }
\label{tab:strategy}
\centering
\begin{tabular}{llcccc}
\hline
Para. & Emo. & V-Prec. & V-F1 & NV-F1 & M-F1 \\ \hline
Freeze & Freeze   & 0.00  & 0.00  & 79.45 & 39.72 \\
Fine T. & Freeze  & 44.90 & 50.87 & 68.16 & 59.52 \\
Freeze & Fine T.  & \textbf{61.90} & 27.08 & \textbf{79.65} & 53.37 \\
LoRA & Fine T.    & 48.53 & 46.15 & 74.07 & 60.11 \\
Fine T. & LoRA    & 48.61 & 47.62 & 73.72 & 60.67 \\
Fine T. & Fine T. & 47.96 & \textbf{54.34} & 70.41 & \textbf{62.37} \\
\hline
\end{tabular}
\end{table}

\section{Conclusion and Future Work}

We proposed a paralinguistic- and emotion-aware model to better capture the correct timing to express validation without relying on dialogue context. The experiments demonstrate that incorporating both paralinguistic and emotional cues enables the model to detect appropriate validation timing more accurately. The proposed method achieved a validation-class precision and F1 of 47.96\% and 54.34\%, with absolute gains of 10.92\% and 6.72\% over current speech baselines, and 6.64\% and 9.69\% while compared to current text baselines. In future work, we will extend the framework to multimodal and multilingual settings and deploy it on an android robot platform for real-world human–robot interaction.


\section{Acknowledgment}
This work was supported by JST NEXUS 251043539, and JST Moonshot R\&D JPMJPS2011.

\bibliographystyle{IEEEbib}
\bibliography{strings,refs}

\end{document}